%% file: main.tex
\documentclass[parskip]{scrartcl}

\usepackage{amsmath,amssymb,amsthm}
\usepackage{bbm}
\usepackage{hyperref}

\newtheorem{theorem}{Theorem}[section]
\newtheorem{proposition}[theorem]{Proposition}
\newtheorem{lemma}[theorem]{Lemma}

\newcommand{\hilbert}{\mathcal{H}}      % Hilbert space
\newcommand{\fock}{\mathcal{F}}         % Fock space

\newcommand{\N}{\mathbbm{N}}

\newcommand{\R}{\mathbbm{R}}

\newcommand{\id}{\mathbbm{1}}            % Identity operator

\DeclareMathOperator{\mathspan}{span}    % Linear hull
\begin{document}
\title{Convergence of Infinite Series in Bosonic Second Quantization}
\author{Peter Otte \\ Fakult\"at f\"ur Mathematik \\ Ruhr-Universit\"at Bochum \\
Germany}
%\date{}
%\subjclass[2010]{81S05 47N50 81R15}
\maketitle
\begin{abstract}
\input{abstract}
\end{abstract}
\section{Introduction\label{introduction}}
\input{introduction}
\section{The CCR and Second Quantization\label{ccr}}
\input{ccr}
\section{Prerequisites from Operator Theory\label{prerequisites}}
\input{prerequisites}
\section{The Number Operator\label{number}}
\input{number}
\section{General Quadratic Operators\label{operators}}
\input{operators}
\section*{Acknowledgment}
\input{acknowledgment}
\bibliographystyle{plain}
\bibliography{literature}
\end{document}

%% file: abstract.tex
The functor of second quantization as well as quadratic creation and
annihilation operators on the bosonic Fock space are defined through possibly
infinite series. The domain of convergence is investigated by precise number
operator estimates and the Banach-Steinhaus Theorem. Some fundamental properties
of the limit operators are derived.

%% file: introduction.tex
The functor of second quantization is one of the so-called quadratic operators,
which are used to quantize one-particle Hamiltonians, to construct
representations of certain Lie algebras, and so on. There are three types of
such operators, to wit
\begin{equation}\label{intro_operators}
\begin{gathered}
  d\Gamma(B)  := \sum_j a^\dagger(Be_j)a(\bar e_j), \\
  \Delta(A)   := \sum_j a(Ae_j)a(\bar e_j),\ 
  \Delta^+(C) := \sum_j a^\dagger(Ce_j)a^\dagger(\bar e_j) .
\end{gathered}
\end{equation}
Here, we are working in a Fock representation of the canonical commutation
relations (CCR) over a separable complex Hilbert space $L$ with $a(f)$ and
$a^\dagger(f)$ being the usual annihilation and creation operators on the Fock
space $\fock$. $A$, $B$, $C$ are some linear operators on $L$ and $\{e_j\}$ is
a complete orthonormal system (ONS) in $L$. Section \ref{ccr} provides details
and background.

When $\dim L=\infty$ the sums in \eqref{intro_operators} are infinite series
and, therefore, we need to look into convergence. Theorems \ref{B}, \ref{A},
\ref{C} say that the series converge strongly on some dense $D_1\subset\fock$
which is what one can expect at best given that $a(f)$ and $a^\dagger(f)$ are
unbounded. It turns out that $d\Gamma(B)$ is well-defined for bounded $B$
whereas $\Delta(A)$ and $\Delta^+(C)$ need Hilbert-Schmidt operators $A$,
$C$. In some sense these conditions are necessary, see Propositions
\ref{B_necessary}, \ref{A_necessary}, \ref{C_necessary}.

The proofs are based upon the Banach-Steinhaus Theorem
\ref{banach_steinhaus}, which actually deals with bounded operators. To make our
operators bounded we define norms with the aid of the number operator
$N:=d\Gamma(\id)$, which is studied thoroughly in Section \ref{number}. The
difficult part then is to check the uniform boundedness condition in the
Banach-Steinhaus Theorem. To this end, we derive precise number operator
estimates for the partial sums of $d\Gamma(B)$, $\Delta(A)$, $\Delta^+(C)$ in
Lemmas \ref{B_bound}, \ref{A_bound}, \ref{C_bound} being true for the limit
operators as well. The bounds for $\Delta(A)$ and $\Delta^+(C)$ improve those
previously derived in
\cite[(2.1), (2.2)]{BruneauDerezinski2007},\cite[(2.3), (2.4)]{HiroshimaIto2004},
\cite[Lemma 3.6]{HoneggerRieckers1996}, \cite[(70)]{GrosseLangmann1992}.

All results can be transferred to fermionic operators, i.e. operators satisfying
the canonical anticommutation relations instead. In that case, all number
operator estimates can be strengthened provided the operators $A$, $B$, $C$
belong to some von Neumann-Schatten class, see \cite{Otte2010}.

%% file: ccr.tex
We will be working against the background of bosonic Fock space theory laid down
in axioms \eqref{ccr_domain} through \eqref{cyclicity} below (see also
\cite{Ottesen1995}, \cite{CareyRuijsenaars1987}, \cite{KristensenMejlboPoulsen1965}).

Let $L$ be a complex Hilbert space. For our purposes it is reasonable to require
$L$ to be separable with $\dim L=\infty$, without the latter some questions
becoming trivial. Furthermore, there is a conjugation $J:L\to L$, $f\mapsto\bar
f$, compatible with the scalar product. By $B_\infty(L)$ and $B_p(L)$, $p\geq
1$, we denote the bounded operators and the von Neumann-Schatten classes on $L$,
respectively. Finally, for $A\in B_\infty(L)$ we define the conjugate $\bar
A:=JAJ$ and the transpose $A^T:=\bar A^*$.

Let $\fock$ be another complex Hilbert space. We take $L$ as index space for
operators in $\fock$, i.e. an operator-valued functional is a map $f\mapsto
c(f)$ where $c(f)$ is an operator in $\fock$ depending linearly on $f$. The CCR
are concerned with two such functionals $a$, $a^\dagger$. We assume there
is a common dense domain of definition $D\subset\fock$ with
\begin{equation}\label{ccr_domain}
  a(f)D\subset D,\ a^\dagger(f)D\subset D\ \text{for all}\ f\in L.
\end{equation}
These operators are said to give a representation of the CCR if for all 
$f,g\in L$ on $D$
\begin{gather}
  [a(f),a(g)] = 0 = [a^\dagger(f),a^\dagger(g)], \label{ccr1} \\
  [a(f),a^\dagger(g)] = (\bar f, g)\id \label{ccr2}
\end{gather}
where the square brackets denote the commutator. We further require the unitarity
condition, roughly saying $a(f)^* = a^\dagger(\bar f)$. More precisely,
\begin{equation}\label{unitarity}
  (a(f)\Phi,\Psi)  = (\Phi,a^\dagger(\bar f)\Psi) \ \text{for all}\
  \Phi,\Psi\in D .
\end{equation}
A Fock representation has a vacuum $\Omega\in\fock$, $\|\Omega\|=1$, that
is annihilated by the $a(f)$'s
\begin{equation}\label{vacuum}
  a(f)\Omega = 0\ \text{for all}\ f\in L
\end{equation}
and cyclic for the $a^\dagger(f)$'s, i.e. 
\begin{equation}\label{cyclicity}
  \fock = \bar\fock_0,\
  \fock_0 := \mathspan\{a^\dagger(f_{j_n})\cdots a^\dagger(f_{j_1})\Omega\mid
  n\in\N_0\} .
\end{equation}
Consequently, $a(f)$ and $a^\dagger(f)$ are called annihilation and creation
operator, respectively. Since most of the explicit calculations are carried out
on $\fock_0$ we will use the notation
\begin{equation*}
  a^\dagger(f_n)\cdots \widehat{a^\dagger(f_j)} \cdots a^\dagger(f_1)
    := a^\dagger(f_n)\cdots a^\dagger(f_{j+1})a^\dagger(f_{j-1}) \cdots a^\dagger(f_1)
\end{equation*}
to indicate that a factor is missing. The structure of a Fock representation is
fairly detailed determined by the axioms.

\begin{theorem}\label{fock_space}
(a) The $n$-particle spaces $\fock^{(n)}$
\begin{equation}\label{n_particle_space}
  \fock^{(n)}:=\bar \fock_0^{(n)},\ 
  \fock_0^{(n)} := \mathspan\{ a^\dagger(f_n)\cdots
  a^\dagger(f_1)\Omega\},\ 
   n\geq 0,
\end{equation}
are orthogonal to each other. The subspace of finite particle numbers
\begin{equation}\label{F0}
  \fock_\mathit{fin} := \mathspan\{ \Phi \mid \Phi\in\fock^{(n)},\ n\in\N_0\}
\end{equation}
is dense in the Fock space, $\fock = \bar\fock_\mathit{fin}$, i.e. $\fock$
is the orthogonal sum of the $n$-particle spaces in the Hilbert space sense
\begin{equation*}
  \fock = \bigoplus_{n=0}^\infty \fock^{(n)},\
  \Phi = \Phi^{(0)} + \Phi^{(1)} + \Phi^{(2)} +\cdots ,\
  \Phi^{(n)}\in\fock^{(n)} .
\end{equation*}
(b) $\fock^{(n)}$ and the symmetric tensor product $L^{\otimes n}$ are
isomorphic Hilbert spaces with the scalar products related via
\begin{equation}\label{scalar_product}
  (a^\dagger(f_n)\cdots a^\dagger(f_1)\Omega,a^\dagger(g_n)\cdots a^\dagger(g_1)\Omega)
    = \sum_{\sigma\in S_n}\prod_{j=1}^n(f_j,g_{\sigma(j)})
\end{equation}
where $S_n$ are all permutations of $\{1,\ldots,n\}$.
\end{theorem}

For the creation and annihilation operators it is obvious
\begin{equation*}
  a^\dagger(f):\fock_0^{(n)}\to\fock_0^{(n+1)} ,\ 
  a(f) : \fock_0^{(n)}\to\fock_0^{(n-1)} .
\end{equation*}
Furthermore, they are closable operators due to the unitarity condition
\eqref{unitarity} and $D$ being dense. By the Wielandt-Wintner theorem
\cite{Wielandt1949}, \cite{Wintner1947} they cannot be bounded on
$\fock$. However, they are bounded on the $n$-particle spaces.

\begin{theorem}\label{ccr_continuity}
The operators $a(f)$ and $a^\dagger(f)$ can be
extended to $\fock^{(n)}$ and, thus, to $\fock_\mathit{fin}$. The extended operators
satisfy the CCR and
\begin{equation}\label{ccr_continuity01}
  \sup_{\substack{\Phi\in\fock^{(n)} \\ \|\Phi\|=1}}
    \|a(f)\Phi\| = \sqrt{n} \|f\| ,\  
  \sup_{\substack{\Phi\in\fock^{(n)} \\ \|\Phi\|=1}}
    \|a^\dagger(f)\Phi\| = \sqrt{n+1} \|f\|
      ,\ f\in L,\ n\in\N_0 .
\end{equation}
\end{theorem}

Motivated by \eqref{ccr_continuity01} we single out certain subspaces of $\fock$.

\begin{theorem}\label{norm}
Let $\Phi\in\fock$, $\Phi=\Phi^{(0)}+\Phi^{(1)}+\cdots$ with
$\Phi^{(n)}\in\fock^{(n)}$, and $\alpha\geq 0$. Then,
\begin{equation*}
  D_\alpha := \{\Phi\in\fock\mid 
    \sum_{n=0}^\infty (n+1)^{2\alpha} \|\Phi^{(n)}\|^2<\infty\},\
  \|\Phi\|_\alpha^2 := \sum_{n=0}^\infty (n+1)^{2\alpha} \|\Phi^{(n)}\|^2
\end{equation*}
is a Hilbert space with norm $\|\cdot\|_\alpha$ the related properties having
the prefix $\alpha$. 
The subspaces $\fock_0,\fock_\mathit{fin}\subset D_\alpha$ are $\alpha$-dense and so
is $D_\beta\subset D_\alpha$ for $\beta\geq\alpha$.
\end{theorem}

\begin{theorem}\label{ccr_final}
The operators $a(f)$ and $a^\dagger(f)$ can be extended to
$D_{1/2}$ with
\begin{equation*}
  \|a(f)\Phi\|\leq\|f\| \|\Phi\|_{1/2},\
  \|a^\dagger(f)\Phi\| \leq \|f\| \|\Phi\|_{1/2} .
\end{equation*}
In particular, the maps $f\mapsto a(f)\Phi$ and $f\mapsto a^\dagger(f)\Phi$ are
continuous for fixed $\Phi\in D_{1/2}$. Furthermore, $a(f)$ and $a^\dagger(f)$
satisfy the CCR on $D_1$.
\end{theorem}

%% file: prerequisites.tex
We collect the necessary prerequisites starting with the Banach-Steinhaus
Theorem (BST).

\begin{theorem}[Banach-Steinhaus]\label{banach_steinhaus}
Let $X$ and $Y$ be Banach spaces and $(F_n)$ be a sequence of bounded linear
operators $F_n:X\to Y$. Then, $(F_n)$ converges strongly on $X$ if and only if
\begin{enumerate}
\item $(F_n)$ converges strongly on a dense subset $U\subset X$.
\item The $F_n$ are uniformly bounded, $\|F_n\| \leq C$ for all $n\in\N$ and 
some $C$.
\end{enumerate}
Furthermore, $F\varphi:=\lim\limits_{n\to\infty} F_n\varphi$ is bounded with
$\|F\| \leq C$.
\end{theorem}

To check the boundedness required by the BST we need some operator inequalities,
which are to be understood in the sense of quadratic forms. At first, we
generalize Cauchy-Schwarz's inequality.

\begin{lemma}\label{cauchy_schwarz}
Let $\hilbert$ be a Pre-Hilbert space and $a_j,b_j:\hilbert\to\hilbert$
be linear operators with adjoints $a_j^*,b_j^*:\hilbert\to\hilbert$. Then,
\begin{equation*}
  \pm\sum_{j,k=1}^M a_j^*b_k^*b_ja_k
    \leq\sum_{j,k=1}^M a_j^*b_k^*b_ka_j .
\end{equation*}
\end{lemma}
\begin{proof}
The inequality follows from
\begin{equation*}
  0 \leq \sum_{j,k=1}^M (\sigma b_ka_j-b_ja_k)^*(\sigma b_ka_j-b_ja_k) 
    = 2\sum_{j,k=1}^M(a_j^*b_k^*b_ka_j - \sigma a_j^*b_k^*b_ja_k) 
\end{equation*}
where $\sigma\in\{\pm 1\}$. 
\end{proof}

We reformulate some linear algebra within our framework.

\begin{lemma}\label{diagonalization}
Let $a(f)$ and $a^\dagger(f)$ be operator-valued functionals satisfying
\eqref{ccr_domain} and \eqref{unitarity}. Let $A,B\in B_\infty(L)$ and
$\{e_j\}$, $j=1,\ldots,M$, be an ONS in $L$. Then,
\begin{multline}\label{diagonalization01}
  0 \leq \sum_{j=1}^M a^\dagger(BAe_j)a(\bar B\bar A\bar e_j) = \\
    = \sum_{j,k=1}^M (A^* e_j',A^* e_k')a^\dagger(Be_j')a(\bar B\bar e_k')
    \leq \|A\| \sum_{j=1}^M a^\dagger(Be_j')a(\bar B\bar e_j')
\end{multline}
where $\{e_j'\}$ is any ONS such that
$\mathspan\{Ae_1,\ldots,Ae_M\}\subset\mathspan\{e_1',\ldots,e_M'\}$.
\end{lemma}
\begin{proof}
The restriction $A_M$ of $A$ to $\mathspan\{e_1,\ldots,e_M\}$ is compact. Therefore,
\begin{equation*}
  A_M g_j = \mu_j f_j,\ A_M^* f_j = \mu_j g_j
\end{equation*}
with singular vectors and values. Obviously,
\begin{equation*}
\begin{split}
  \sum_{j=1}^M a^\dagger(BA_Mg_j)a(\bar B\bar A_M\bar g_j)
     & = \sum_{j=1}^M \mu_j^2 a^\dagger(Bf_j)a(\bar B\bar f_j) \\
     & = \sum_{j,k=1}^M (A_M^*f_j,A_M^*f_k) a^\dagger(Bf_j)a(\bar B\bar f_k) .
\end{split}
\end{equation*}
To conclude the proof recall the estimate $0\leq\mu_j\leq \|A\|=\|A^*\|$ and note that in the
first and last sum each ONS can be replaced by one that spans the same subspace.
\end{proof}
  
The preceding Lemmas \ref{cauchy_schwarz} and \ref{diagonalization} are
concerned with quadratic forms, which of course can be written as inequalities
involving the norm instead. For unbounded operators, like creation and
annihilation operators, the norm inequalities make sense on a larger domain,
however, without being proven there. This can be overcome by a simple
observation, which becomes essential when treating the partial sums below.

\begin{lemma}\label{extension}
Let $X$, $Y$ be normed spaces and $F:X\to Y$ be a bounded linear
operator. If $\|F\varphi\|\leq c\|\varphi\|$ on a dense subset $U\subset X$ then
$\|F\|\leq c$.
\end{lemma}
\begin{proof}
Approximation argument.
\end{proof}

%% file: number.tex
To define quadratic operators we look at the simplest such operator
$N:=d\Gamma(\id)$, the particle number operator or number operator for short. In
what follows, $\{e_j\}$ is a complete ONS in $L$. We study the partial sums
\begin{equation*}
  N_M := \sum_{j=1}^M a^\dagger(e_j)a(\bar e_j)
\end{equation*}
and work our way up to the maximal domain of convergence.

\begin{lemma}\label{N_convergence}
On $\fock_0$ the strong limit $\lim\limits_{M\to\infty}N_M =: N$
exists and reads
\begin{equation*}
  N\Phi = n\Phi\ \text{for}\ \Phi\in\fock_0^{(n)}.
\end{equation*}
\end{lemma}
\begin{proof}
Take the limit $M\to\infty$ in
\begin{equation*}
    N_M a^\dagger(f_n)\cdots a^\dagger(f_1)\Omega
        = \sum_{k=1}^n \sum_{j=1}^M (e_j,f_k)a^\dagger(e_j)a^\dagger(f_n)\cdots 
           \widehat{a^\dagger(f_k)}\cdots a^\dagger(f_1)\Omega
\end{equation*}
and use that $a^\dagger(f)\Phi$ is continuous in $f$ by Theorem
\ref{ccr_continuity}.
\end{proof}

The boundedness condition of the BST \ref{banach_steinhaus} ought to be
intuitively clear.

\begin{lemma}\label{N_bound}
The operator $N_M$ satisfies
\begin{equation*}
  \|N_M\Phi\|\leq \|\Phi\|_1 \ \text{for}\ \Phi\in D_1 .
\end{equation*}
\end{lemma}
\begin{proof}
It is enough to show the estimate on each $\fock^{(n)}$. For $n=0$ it is
trivial. Assume it is true for some $n\geq 0$. When $\Phi\in\fock^{(n+1)}$,
\begin{equation*}
\begin{split}
  \| N_M\Phi\|^2
      & = \sum_{j=1}^M(a(\bar e_j)\Phi, N_Ma(\bar e_j)\Phi) +
               (\Phi,N_M\Phi) \\
      & \leq \sum_{j=1}^M \|a(\bar e_j)\Phi\| \|N_M a(\bar f)\Phi\| 
                + (\Phi,N_M\Phi)
\end{split}
\end{equation*}
since $a(f)\Phi\in\fock^{(n)}$. Note, trivially $\fock^{(n)}\subset
D_{3/2}$. Now,
\begin{equation*}
  \| N_M\Phi\|^2
      \leq n \sum_{j=1}^M \|a(\bar e_j)\Phi\|^2
                + (\Phi,N_M\Phi)
      = (n+1) (\Phi, N_M\Phi)
      \leq (n+1)\|\Phi\| \|N_M\Phi\|
\end{equation*}
completes the proof.
\end{proof}

\begin{theorem}\label{N}
The $N_M$ converge strongly on $D_1$ to a well-defined operator $N$ with
\begin{equation}\label{N01}
  (N\Phi)^{(n)} = n\Phi^{(n)} .
\end{equation}
\end{theorem}
\begin{proof}
The $N_M$ converge strongly on the $1$-dense subset $\fock_0\subset
D_1$, Lemma \ref{N_convergence}, and their norms are bounded by Lemma
\ref{N_bound}. Hence, by the BST \ref{banach_steinhaus}
the $N_M$ converge on the whole $D_1$ to a $1$-bounded operator.
The partial sums $\tilde N_M$ formed with another ONS will converge
as well to, say, $\tilde N$. By Lemma \ref{N_convergence} on $\fock_0$ the limit
looks the same whatever ONS we choose. Hence, the $1$-bounded operators $N$
and $\tilde N$ coincide on the $1$-dense subset $\fock_0$ and, thus,
everywhere on $D_1$.
\end{proof}

We note some properties of $N$, in particular we relate the $\alpha$-norms to $N$.

\begin{theorem}\label{N_properties}
$N^\alpha$ is non-negative and self-adjoint on $D_\alpha$ for $\alpha\geq 0$.
Furthermore, $\|\Phi\|_\alpha = \|(N+\id)^\alpha\Phi\|$ for $\Phi\in D_\alpha$.
On $D_{3/2}$,
\begin{equation}\label{N_properties01}
  [ N, a(f) ] = -a(f),\ [N, a^\dagger(f)] = a^\dagger(f).
\end{equation}
\end{theorem}
\begin{proof}
The spectral decomposition of $N$ is given by \eqref{N01} which implies $N$ is
non-negative and self-adjoint and also gives a representation for
$N^\alpha$. The norm relation is obvious from Definition \ref{norm}. On
$D_{3/2}$
\begin{multline*}
  [ \sum_{j=1}^M a^\dagger(e_j)a(\bar e_j), a(f)] 
     = \sum_{j=1}^M [a^\dagger(e_j),a(f)] a(\bar e_j) = \\
     = - \sum_{j=1}^M (\bar f,e_j) a(\bar e_j)
     = - a( \sum_{j=1}^M (\bar e_j,f) \bar e_j) .
\end{multline*}
The leftmost term converges strongly on $D_{3/2}$ to $[N,a(f)]$ and the
rightmost term to $-a(f)$ since $a(f)\Phi$ is continuous in $f$. The second
relation follows in like manner.
\end{proof}

Because Lemma \ref{N_convergence} needs a vacuum so does Theorem \ref{N}.
However, there are representations other than the Fock representation that have
a number operator when it is defined through strongly converging partial
sums. There again, it can satisfy the commutation relations
\eqref{N_properties01} only in a Fock representation. For a thorough discussion
see \cite{Chaiken1967}, \cite{Chaiken1968}, \cite{DellAntonioDoplicher1967},
\cite{DellAntonioDoplicherRuelle1966}.

To treat general quadratic operators we need some technical estimates. To begin
with, we rewrite Lemma \ref{diagonalization} in a nearly obvious way.

\begin{lemma}\label{basic_estimate}
Let $A\in B_\infty(L)$. Then, on $D_1$
\begin{equation*}
  \sum_{j=1}^M a^\dagger(Ae_j)a(\bar A \bar e_j) \leq \|A\|^2 N .
\end{equation*}
\end{lemma}
\begin{proof}
Apply \eqref{diagonalization01} and bound the $N_M$ by $N$.
\end{proof}

It is instructive to prove the next result with the aid of the commutators
\eqref{N_properties01} instead of \eqref{N01}.

\begin{lemma}\label{technical_estimate}
Let $A\in B_\infty(L)$, $\{e_j\}$ be an ONS in $L$, and $f_j:=Ae_j$. Then, on $D_2$
\begin{equation*}
  \sum_{j=1}^M a^\dagger(f_j)(N+\id) a(\bar f_j) 
     \leq \|A\|^2 N^2
  \ \text{and}\
  \sum_{j=1}^M a^\dagger(e_j)Na(\bar e_j) \leq N(N-\id) .
\end{equation*}
\end{lemma}
\begin{proof}
With the commutator \eqref{N_properties01} and $\gamma\in\R\setminus\{0\}$ on $D_2$,
\begin{equation*}
\begin{split}
  2\sum_{j=1}^M a^\dagger(f_j)N a(\bar f_j) 
     & = -\Big(\frac{1}{\gamma}\sum_{j=1}^M 
            a^\dagger(f_j)a(\bar f_j)-\gamma N\Big)^2  + \gamma^2 N^2 \\
     & \quad  +\frac{1}{\gamma^2}\sum_{j,k=1}^M
            a^\dagger(f_j)a^\dagger(f_k)a(\bar f_k)a(\bar f_j) \\
     & \quad +\frac{1}{\gamma^2}\sum_{j,k=1}^M (f_j,f_k)a^\dagger(f_j)a(\bar f_k)
          - 2\sum_{j=1}^M a^\dagger(f_j)a(\bar f_j) .
\end{split}
\end{equation*}
We drop the first term and estimate the quartic term and the first quadratic
term via \eqref{diagonalization01}. To prove the first estimate choose
$\gamma=\|A\|$, the case $\|A\|=0$ being trivial.
To prove the second note for every $M'\geq M$ 
\begin{equation*}
  \sum_{j=1}^M a^\dagger(e_j)Na(\bar e_j)
   \leq  \sum_{j=1}^{M'} a^\dagger(e_j)Na(\bar e_j)
      \leq N^2 - \sum_{j=1}^{M'} a^\dagger(e_j)a(\bar e_j)
\end{equation*}
by the first part. To conclude, let $M'\to\infty$ and note the strong
convergence on $D_1$.
\end{proof}

%% file: operators.tex
We are going to study the general quadratic operators from
\eqref{intro_operators} by means of the respective partial sums
\begin{gather*}
  d\Gamma_M(B) := \sum_{j=1}^M a^\dagger(Be_j)a(\bar e_j), \\
  \Delta_M(A) :=  \sum_{j=1}^M a(Ae_j)a(\bar e_j),\
  \Delta^+_M(C) := \sum_{j=1}^M a^\dagger(Ce_j)a^\dagger(\bar e_j) .
\end{gather*}
Hereinafter, $\{e_j\}$ always is a complete ONS in $L$ and $A,B,C\in B_\infty(L)$. 
We will treat the three cases separately since the calculations
differ in some points. However, the overall strategy is the same: 
The partial sums converge on $\fock_0$ and can be bounded with the aid of $N$.
Then, the BST \ref{banach_steinhaus} extends convergence to $D_1$. 
We start with $d\Gamma(B)$.

\begin{lemma}[Convergence]\label{B_convergence}
For $B\in B_\infty$ the strong limit
$\lim\limits_{M\to\infty}d\Gamma_M(B)=:d\Gamma(B)$ exists on $\fock_0$ with
\begin{equation}\label{B_convergence01}
  d\Gamma(B)a^\dagger(f_n)\cdots a^\dagger(f_1)\Omega =
    \sum_{j=1}^n a^\dagger(f_n)\cdots a^\dagger(Bf_k)\cdots
  a^\dagger(f_1)\Omega,\ n\geq 1,
\end{equation}
and $d\Gamma(B)\Omega = 0$.
\end{lemma}
\begin{proof}
Follows from
\begin{equation}\label{B_convergence02}
  a^\dagger(Be_j)a(\bar e_j)a^\dagger(f_n)\cdots a^\dagger(f_1)\Omega
     = \sum_{k=1}^n (e_j,f_k)a^\dagger(Be_j)a^\dagger(f_n)\cdots
          \widehat{a^\dagger(f_k)} \cdots a^\dagger(f_1)\Omega
\end{equation}
since $f\mapsto a^\dagger(f)\Phi$ is continuous for fixed $\Phi$ and $B$ is
bounded.
\end{proof}

\begin{lemma}[Boundedness]\label{B_bound}
Let $B\in B_\infty(L)$. Then, for $\Phi\in D_1$
\begin{equation*}
  \|d\Gamma_M(B)\Phi\| \leq \|B\| \|N\Phi\|.
\end{equation*}
\end{lemma}
\begin{proof}
After normal ordering we have on $D_2$
\begin{equation*}
\begin{split}
  d\Gamma_M(B)^*d\Gamma_M(B)
    & = \sum_{j,k=1}^M a^\dagger(e_j)a^\dagger(Be_k)a(\overline{Be_j})a(\bar e_k)
          + \sum_{j,k=1}^M (Be_j,Be_k)a^\dagger(e_j)a(\bar e_k) \\
    & \leq \|B\|^2\sum_{j=1}^M a^\dagger(e_j) N a(\bar e_j) +
            \|B\|^2 \sum_{j=1}^Ma^\dagger(e_j)a(\bar e_j)
\end{split}
\end{equation*}
where we used the Cauchy-Schwarz inequality \ref{cauchy_schwarz} and then Lemma
\ref{diagonalization}.
Now, Lemma \ref{technical_estimate} proves the estimate on $D_2$. To extend
it we use the crude estimate on $D_1$
\begin{equation*}
  \|d\Gamma_M(B)\Phi\|
    \leq \sum_{j=1}^M \|a^\dagger(Be_j)a(\bar e_j)\Phi\|
    \leq 2 \sum_{j=1}^M \|Be_j\| \|N\Phi\| .
\end{equation*}
Hence, Lemma \ref{extension} completes the proof.
\end{proof}

\begin{theorem}\label{B}
For $B\in B_\infty(L)$ the partial sums $d\Gamma_M(B)$ converge strongly on
$D_1$ to a well-defined operator $d\Gamma(B)$ with
\begin{equation*}
  \|d\Gamma(B)\Phi\|\leq\|B\|\|N\Phi\|,\ \Phi\in D_1 .
\end{equation*}
Furthermore, $d\Gamma(B)^* = d\Gamma(B^*)$ on $D_1$.
\end{theorem}
\begin{proof}
By Lemma \ref{B_convergence} the partial sums $d\Gamma_M(B)$ converge on
$\fock_0$. Along with the bound from Lemma \ref{B_bound} the BST
\ref{banach_steinhaus} implies strong convergence on $D_1$ and the 
bound for $d\Gamma(B)$.
If $d\tilde\Gamma(B)$ is the limit obtained via another ONS it is an
$1$-bounded operator as well as $d\Gamma(B)$. By \eqref{B_convergence01} they
coincide on the $1$-dense subset $\fock_0$ and, thus, everywhere on
$D_1$. Likewise, using \eqref{B_convergence01} and \eqref{scalar_product} one
concludes that for $\Phi,\Psi\in\fock_0$
\begin{equation*}
  (\Phi, d\Gamma(B)\Psi) = (d\Gamma(B^*)\Phi,\Psi)
\end{equation*}
which extends to $D_1$. Thus, $d\Gamma(B)^*=d\Gamma(B^*)$ on $D_1$.
\end{proof}

In a way, $B$ being bounded is necessary when we want to have convergence on
all of $\fock^{(1)}$. If we drop this we can of course second quantize
unbounded operators as well. 

\begin{proposition}\label{B_necessary}
If $B$ is defined on $L_0:=\mathspan\{e_j\mid j\in\N\}$ and the corresponding
$d\Gamma_M(B)$ converge strongly on all of $\fock^{(1)}$ then
$B\in B_\infty(L_0)$.
\end{proposition}
\begin{proof}
Strong convergence on $\fock^{(1)}$ implies for all $M$
\begin{equation*}
  \|d\Gamma_M(B) \Phi\| \leq c \|\Phi\|,\ \Phi\in\fock^{(1)},
\end{equation*}
by the BST \ref{banach_steinhaus} with some constant $c$. From this, \eqref{B_convergence02} with
$n=1$, and \eqref{scalar_product} we get
\begin{multline*}
  c \|f\|
     = c\|a^\dagger(f)\Omega\| 
   \geq \| d\Gamma_M(B)a^\dagger(f)\Omega\|  = \\
     = \| a^\dagger(B\sum_{j=1}^M (e_j,f)e_j)\Omega \|
     = \|B \sum_{j=1}^M (e_j,f)e_j\| .
\end{multline*}
When $f\in L_0$ choose $M$ large enough and conclude that $B$ is bounded on
$L_0$.
\end{proof}

The $\Delta_M(A)$ converge, initially, for a much larger class of operators $A$.

\begin{lemma}[Convergence]\label{A_convergence}
Let $A\in B_\infty(L)$. The strong limit
$\lim\limits_{M\to\infty}\Delta_M(A)=:\Delta(A)$ exists on $\fock_0$ and reads
\begin{equation}\label{A_convergence01}
     \Delta(A)
     a^\dagger(g_n)\cdots a^\dagger(g_1)\Omega  
     = \sum_{\substack{k,l=1 \\ k\neq l}}^n 
       (\bar g_l,Ag_k)a^\dagger(g_n)\cdots\widehat{a^\dagger(g_l)}
         \cdots\widehat{a^\dagger(g_k)}\cdots a^\dagger(g_1)\Omega .
\end{equation}
\end{lemma}
\begin{proof}
Follows immediately from
\begin{multline*}
  a(Ae_j)a(\bar e_j)a^\dagger(g_n)\cdots a^\dagger(g_1)\Omega \\
     =  \sum_{\substack{k,l=1\\l\neq k}}^n 
             (e_j,g_k)(\overline{Ae_j},g_l)
               a^\dagger(g_n)\cdots\widehat{a^\dagger(g_l)}
                \cdots\widehat{a^\dagger(g_k)}\cdots a^\dagger(g_1)\Omega ,
\end{multline*}
the continuity of the scalar product, and the boundedness of $A$.
\end{proof}

\begin{lemma}[Boundedness]\label{A_bound} 
Let $A\in B_2(L)$. Then on $D_1$,
\begin{equation*}
  \|\Delta_M(A)\Phi\|^2
      \leq \|A\|^2 \|N\Phi\|^2 + (\|A\|_2^2-\|A\|^2)\|N^{1/2}\Phi\|^2 .
\end{equation*}
\end{lemma}
\begin{proof}
Let us write $f_j:=Ae_j$ for short. On $D_2$,
\begin{equation*}
  \Delta_M(A)^*\Delta_M(A)
      = \sum_{j,k=1}^M a^\dagger(e_j)a(f_k)a^\dagger(\bar f_j)a(\bar e_k)
             - \sum_{j,k=1}^M (\bar f_k,\bar f_j) a^\dagger(e_j)a(\bar e_k) .
\end{equation*}
The rightmost sum is positive by Lemma \ref{diagonalization} and may be dropped.
By Cauchy-Schwarz \ref{cauchy_schwarz} and Lemmas \ref{diagonalization} and
\ref{technical_estimate}
\begin{equation*}
\begin{split}
\Delta_M(A)^*\Delta_M(A)
     & \leq \sum_{j,k=1}^M a^\dagger(e_j)a^\dagger(\bar f_k)a(f_k)a(\bar e_j)
          +\sum_{j,k=1}^M (\bar f_k,\bar f_k) a^\dagger(e_j)a(\bar e_j) \\
     & \leq \|A\|^2 \sum_{j=1}^M a^\dagger(e_j)Na(\bar e_j) + \|A\|_2^2 N \\
     & \leq \|A\|^2 N(N-1) + \|A\|_2^2 N .
\end{split}
\end{equation*}
This proves the estimate on $D_2$. On $D_1$ we have
\begin{equation*}
  \|\Delta_M(A)\Phi\|
    \leq \sum_{j=1}^M \|Ae_j\| \|N\Phi\|
\end{equation*}
which allows us by dint of Lemma \ref{extension} to extend the estimate to $D_1$.
\end{proof}

The Hilbert-Schmidt condition, characteristic of the operator $\Delta(A)$ as
well as for $\Delta^+(C)$ below, originates from this Lemma \ref{A_bound}.

\begin{theorem}\label{A}
Let $A\in B_2(L)$. Then on $D_1$ the partial sums $\Delta_M(A)$ converge
strongly to a well-defined operator $\Delta(A)$ with
\begin{equation*}
  \|\Delta(A)\Phi\|^2
   \leq \|A\|^2 \|N\Phi\|^2 + (\|A\|_2^2-\|A\|^2)\|N^{1/2}\Phi\|^2,\ \Phi\in D_1.
\end{equation*}
Furthermore, $\Delta(A^T)=\Delta(A)$. 
\end{theorem}
\begin{proof}
By Lemma \ref{A_convergence} the $\Delta_M(A)$ converge strongly on $\fock_0$. Along
with the bound from Lemma \ref{A_bound} the BST \ref{banach_steinhaus} implies
strong convergence on $D_1$ and the bound for $\Delta(A)$. $\Delta(A)$ being
independent of the ONS follows analogously to Theorem \ref{B} via
\eqref{A_convergence01}. Finally,
\begin{equation*}
  (\bar g_l,Ag_k) = (\bar g_k, A^T g_l)
\end{equation*}
in \eqref{A_convergence01} implies that the $1$-bounded operators $\Delta(A)$
and $\Delta(A^T)$ coincide on the $1$-dense subspace $\fock_0\subset D_1$ and,
thus, on all of $D_1$.
\end{proof}

Even though we could define $\Delta(A)$ on $\fock_0$ for bounded $A$ the
Hilbert-Schmidt condition in Theorem \ref{A} happens to be necessary.

\begin{proposition}\label{A_necessary}
Let $A^T=A$. If $\Delta_M(A)$ converges on $\fock^{(2)}$ then $A\in B_2(L)$.
\end{proposition}
\begin{proof}
Standard calculations give the formula
\begin{equation}\label{A_necessary01}
  \Delta_M(A)\Delta_M(A)^*\Omega 
     = \Big(\sum_{j,k=1}^M |(Ae_j,\bar e_k)|^2 + \sum_{k=1}^M \|Ae_k\|^2\Big)\Omega
     =: \omega_M\Omega .
\end{equation}
Taking both the norm and the scalar product with $\Omega$ yields
\begin{equation*}
  \|\Delta_M(A)\Delta_M(A)^*\Omega\| 
     = \omega_M =   \|\Delta_M(A)^*\Omega\|^2 .
\end{equation*}
Let $A\neq 0$ the case $A=0$ being trivial. Then $\omega_M\neq 0$ for large
$M$. Therefore, we can define unit vectors
\begin{equation*}
  \Phi_M := \frac{1}{\|\Delta_M(A)^*\Omega\|} \Delta_M(A)^*\Omega\in\fock^{(2)}
\end{equation*}
and obtain
\begin{equation*}
  \|\Delta_M(A)\Phi_M\| = \omega_M^{\frac{1}{2}} 
                       \geq \Big(\sum_{j=1}^M\|Ae_k\|^2\Big)^{\frac{1}{2}} .
\end{equation*}
When $A\notin B_2(L)$ the right side will become infinite and so will
the norm of $\Delta_M(A)|_{\fock^{(2)}}$. By the BST \ref{banach_steinhaus} the
$\Delta_M(A)$ cannot converge on the entire Hilbert space $\fock^{(2)}$.
\end{proof}

Unfortunately, one cannot deduce strong convergence of $\Delta_M^+(C)$ from that of
their adjoints $\Delta_M(A)$.

\begin{lemma}[Convergence]\label{C_convergence}
For $C\in B_2(L)$ the strong limit
$\lim\limits_{M\to\infty}\Delta_M^+(C)=:\Delta^+(C)$ exists on $\fock_0$.
\end{lemma}
\begin{proof}
$\Delta_M^+(C)$ and $a^\dagger(f)$ commute.
Since $\Delta_M^+(C)\Omega\in\fock^{(2)}$ and
$a^\dagger(f)|_{\fock^{(n)}}$ is bounded it is enough to show
convergence on $\Omega$. As in \eqref{A_necessary01}, 
\begin{equation}\label{C_convergence02}
  \| \sum_{j=M_1}^{M_2} a^\dagger(Ce_j)a^\dagger(\bar e_j)\Omega\|^2
      = \sum_{j=M_1}^{M_2} \|C e_j\|^2
            +\sum_{j,k=M_1}^{M_2} (\bar e_j,Ce_k)(Ce_j,\bar e_k) .
\end{equation}
Since $C\in B_2(L)$ the right side is a Cauchy sequence and so is
$\Delta_M^+(C)\Omega$.
\end{proof}

\begin{lemma}[Boundedness]\label{C_bound}
Let $C\in B_2(L)$ and $\Phi\in D_1$. Then,
\begin{equation*}
  \| \Delta_M^+(C)\Phi\|^2 
       \leq
    \|C\|^2 \|( N(N+2\id) )^{1/2} \Phi\|^2 
         + \|C\|_2^2 \|(N+2\id)^{1/2}\Phi\|^2 .
\end{equation*}
\end{lemma}
\begin{proof}
Put $f_j:=Ce_j$. On $D_2$, 
\begin{equation*}
  \Delta_M^+(C)^*\Delta_M^+(C)
     = \sum_{j,k=1}^Ma(\bar f_j)a^\dagger(\bar e_k)a(e_j)a^\dagger(f_k)
         + \sum_{j=1}^M a^\dagger(f_j)a(\bar f_j) 
         + \sum_{j=1}^M \|f_j\|^2 \id .
\end{equation*}
With the aid of the Cauchy-Schwarz inequality \ref{cauchy_schwarz}
\begin{equation*}
   \sum_{j,k=1}^M a(\bar f_j)a^\dagger(\bar e_k)a(e_j)a^\dagger(f_k)          
     \leq \sum_{j=1}^M a(\bar f_j)Na^\dagger(f_j) .
\end{equation*}
Normal ordering  and Lemma \ref{technical_estimate} imply the statement for
$D_2$. On $D_1$,
\begin{equation*}
  \|\Delta_M^+(C)\Phi\|
     \leq \sum_{j=1}^M \|Ce_j\| \|(N+\id)\Phi\| .
\end{equation*}
Then, Lemma \ref{extension} shows the statement for $D_1$. 
\end{proof}

\begin{theorem}\label{C}
If $C\in B_2(L)$ the partial sums converge strongly on $D_1$ to a well-defined
operator $\Delta^+(C)$ with
\begin{equation*}
  \|\Delta^+(C)\Phi\|^2\leq \|C\|^2 \|( N(N+2\id) )^{1/2}\Phi\|^2 + \|C\|_2^2
     \|(N+2\id)^{1/2}\Phi\|^2,\ \Phi\in D_1 .
\end{equation*}
Furthermore, $\Delta^+(C^T)=\Delta^+(C)$ and $\Delta^+(C) = \Delta(C^*)^*$ on
$D_1$ when $C^T=C$.
\end{theorem}
\begin{proof}
By Lemmas \ref{C_convergence}, \ref{C_bound} $\Delta_M^+(C)$ converge on
$\fock_0$ and are bounded in such a way that the BST \ref{banach_steinhaus}
applies.
$\Delta^+(C)$ being independent of the ONS follows analogously to Theorem
\ref{B}. Let $\tilde C:=C-C^T=-\tilde C^T$. With the aid of
\eqref{C_convergence02},
\begin{equation*}
  \|(\Delta_M^+(C)-\Delta_M^+(C^T))\Omega\|^2
     = \sum_{j=1}^M \|\tilde Ce_j\|^2 
       - \sum_{j,k=1}^M |(\bar e_j,\tilde Ce_k)|^2
     \to 0,\ M\to\infty,
\end{equation*}
which shows $\Delta^+(C)$ and $\Delta^+(C^T)$ coincide on $\fock_0$ and, thus,
on $D_1$ because of their being $1$-bounded. Finally, when $C^T=C$
\begin{equation*}
  (a(C^*e_j)a(\bar e_j))^* = a^\dagger(e_j)a^\dagger(C^T\bar e_j)
                           = a^\dagger(C\bar e_j)a^\dagger(e_j).
\end{equation*}
This implies the statement for the adjoint since $\Delta_M^+(C)$ and
$\Delta_M(C^*)$ converge independently of the ONS.
\end{proof}

Unlike $\Delta(A)$ the operator $\Delta^+(C)$ needs $C\in B_2(L)$ to exist even
on the vacuum.

\begin{proposition}\label{C_necessary}
Let $C^T=C$. If $\Delta_M^+(C)$ converges on $\Omega$ then $C\in B_2(L)$.
\end{proposition}
\begin{proof}
Follows from \eqref{C_convergence02} where the right side would become infinite
if $C\notin B_2(L)$.
\end{proof}

%% file: acknowledgment.tex
This work was supported by the research network SFB TR 12 -- `Symmetries and
Universality in Mesoscopic Systems' of the German Research Foundation (DFG).

%% file: main.bbl
\begin{thebibliography}{10}

\bibitem{BruneauDerezinski2007}
L.~Bruneau and J.~Derezi{\'n}ski.
\newblock {B}ogoliubov {H}amiltonians and one-parameter groups of {B}ogoliubov
  transformations.
\newblock {\em J. Math. Phys.}, 48(2):022101, 24, 2007.

\bibitem{CareyRuijsenaars1987}
A.~L. Carey and S.~N.~M. Ruijsenaars.
\newblock {O}n fermion gauge groups, current algebras and {K}ac-{M}oody
  algebras.
\newblock {\em Acta Appl. Math.}, 10(1):1--86, 1987.

\bibitem{Chaiken1967}
J.~M. Chaiken.
\newblock {F}inite-particle representations and states of the canonical
  commutation relations.
\newblock {\em Ann. Physics}, 42(1):23--80, 1967.

\bibitem{Chaiken1968}
J.~M. Chaiken.
\newblock {N}umber operators for representations of the canonical commutation
  relations.
\newblock {\em Comm. Math. Phys.}, 8:164--184, 1968.

\bibitem{DellAntonioDoplicher1967}
G.~F. Dell'Antonio and S.~Doplicher.
\newblock {T}otal number of particles and {F}ock representation.
\newblock {\em J. Mathematical Phys.}, 8:663--666, 1967.

\bibitem{DellAntonioDoplicherRuelle1966}
G.~F. Dell'Antonio, S.~Doplicher, and D.~Ruelle.
\newblock {A} theorem on canonical commutation and anticommutation relations.
\newblock {\em Comm. Math. Phys.}, 2:223--230, 1966.

\bibitem{GrosseLangmann1992}
H.~Grosse and E.~Langmann.
\newblock {A} superversion of quasifree second quantization. {I}. {C}harged
  particles.
\newblock {\em J. Math. Phys.}, 33(3):1032--1046, 1992.

\bibitem{HiroshimaIto2004}
F.~Hiroshima and K.~R. Ito.
\newblock {L}ocal exponents and infinitesimal generators of canonical
  transformations on {B}oson {F}ock spaces.
\newblock {\em Infin. Dimens. Anal. Quantum Probab. Relat. Top.},
  7(4):547--571, 2004.

\bibitem{HoneggerRieckers1996}
R.~Honegger and A.~Rieckers.
\newblock {S}queezing {B}ogoliubov transformations on the infinite mode
  {CCR}-algebra.
\newblock {\em J. Math. Phys.}, 37(9):4292--4309, 1996.

\bibitem{KristensenMejlboPoulsen1965}
P.~Kristensen, L.~Mejlbo, and E.~Thue Poulsen.
\newblock {T}empered distributions in infinitely many dimensions. {I}.
  {C}anonical field operators.
\newblock {\em Commun. Math. Phys.}, 1(3):175--214, September 1965.

\bibitem{Otte2010}
P.~Otte.
\newblock {B}oundedness properties of fermionic operators.
\newblock {\em J. Math. Phys.}, 51(8):083503, 12, 2010.

\bibitem{Ottesen1995}
J.~T. Ottesen.
\newblock {\em {I}nfinite-dimensional groups and algebras in quantum physics},
  volume~27 of {\em Lecture Notes in Physics. New Series m: Monographs}.
\newblock Springer-Verlag, Berlin, 1995.

\bibitem{Wielandt1949}
H.~Wielandt.
\newblock \"{U}ber die {U}nbeschr\"anktheit der {O}peratoren der
  {Q}uantenmechanik.
\newblock {\em Math. Ann.}, 121:21, 1949.

\bibitem{Wintner1947}
A.~Wintner.
\newblock {T}he unboundedness of quantum-mechanical matrices.
\newblock {\em Physical Rev. (2)}, 71(10):738--739, 1947.

\end{thebibliography}
